# Photonic Supercoupling in Silicon Topological Waveguides

Ridong Jia[1,2], Yi Ji Tan[1,2], Nikhil Navaratna[1,2], Abhishek Kumar[1,2], Ranjan Singh[1,2,*]

[1]*Division of Physics and Applied Physics, School of Physical and Mathematical Sciences, Nanyang Technological University, 21 Nanyang Link, Singapore 637371, Singapore*

[2]*Centre for Disruptive Photonic Technologies, The Photonics Institute, Nanyang Technological University, 50 Nanyang Avenue, Singapore 639798, Singapore*

[*]Email: ranjans@ntu.edu.sg

Electromagnetic wave coupling between photonic systems relies on the evanescent field typically confined within a single wavelength. Extending evanescent coupling distance requires low refractive index contrast and perfect momentum matching for achieving a large coupling ratio. Here, we report the discovery of photonic supercoupling in a topological valley Hall pair of waveguides, showing a substantial improvement in coupling efficiency across multiple wavelengths. Experimentally, we realize ultra-high coupling ratios between waveguides through valley-conserved vortex flow of electromagnetic energy, attaining 95% coupling efficiency for separations of up to three wavelengths. This demonstration of photonic supercoupling in topological systems significantly extends the coupling distance between on-chip waveguides and components, paving the path for the development of supercoupled photonic integrated devices, optical sensing, and telecommunications.

**Introduction**

Direct coupling between photonic modes typically relies on evanescent wave, which decays exponentially with distance from the mode edge to the bulk space. By strategically constructing and tailoring the alignment, efficient and tunable evanescent coupling can be achieved. This compact coupling technique plays a crucial role in numerous photonic systems such as fiber couplers[1,2], electromagnetic sensors[3–5], microscopies[6], and integrated photonic devices[7,8]. However, the challenge arises when the spatial separation between mode profiles exceeds the evanescent distance, especially when it surpasses a wavelength in free space ($\lambda$). Beyond this limit, the interaction between photonic modes weakens significantly, constraining the inter-system distance to an extremely narrow region.

A long evanescent tail can be achieved for a confined mode when there is a narrow difference in refractive index of the bulk space, potentially facilitating long-distant coupling. For a waveguide mode as an example, the evanescent field can extend over tens of wavelengths when the refractive index contrast between the core and cladding areas is on the order of $10^{-2}$ or even lower. However, achieving such precise control of the indices presents significant challenges across the entire electromagnetic spectrum, particularly in the context of material properties for waveguiding networks and integrated circuit designs, although can be realized in specific scenarios such as the femtosecond laser directly written waveguide arrays[9]. Furthermore, while the evanescent field can encompass the coupled photonic system, the coupling efficiency is highly dependent on the momentum matching condition. Especially in regions beyond $\lambda$, the mode experiences rapid decay such that efficient coupling is further challenged by the sensitivity to momentum disruption. Therefore, achieving efficient coupling over multiple wavelengths pose significant challenges beyond the scope of theoretical models.

Manipulating electromagnetic wave at wavelength- or sub-wavelength-scales offers the potentials to extend the evanescent coupling regions when maintaining coupling efficiency. The construction of spatial inhomogeneities enables precise modulations of the mode profile and momentum, that can be accomplished through the utilization of periodic structures like photonic crystals[10,11]. The extension of mode tails and manipulation of momenta hold the potential to address the limitations of short evanescent coupling, eliminating the dependence on perturbative defect modes[12] or external couplers[13,14]. However, a practical long-range coupling method is still lacking due to the complexities associated with constructing an intermediary medium, capable of conserving the momentum and maintaining corresponding high energy transfer efficiency over distances beyond wavelengths.

Here, we propose and demonstrate a *photonic supercoupling* method in topological valley Hall system[15] designed to extend the coupling distance while achieving a high coupling efficiency. The valley-conserved momentum and the sub-wavelength valley vortex created by breaking spatial inversion symmetry facilitate long-range wave transition within the topological photonic bandgap. We exemplify its application in a waveguide coupling system on a valley photonic crystal (VPC[16,17]) slab, as depicted in Fig. 1(a). Even when separated by multiple wavelengths, the ultra-high waveguide coupling efficiency can be achieved via the structured valley-conserved vortices in the coupling region with $C_{3v}$ point group symmetry. The valley conservation property of the coupling region ensures efficient energy accumulation in the coupled topological waveguide, with an anticipated coupling ratio approaching 100%. The experimental results present an over 95% coupling ratio in a coupling distance of 2.7$\lambda$ and a 31% coupling ratio even at a remarkable 5$\lambda$ distance. The supercoupling technique introduces a new coupling methodology that can enhance the performances of various on-chip topological integrated circuits[18], including topological cavities[19] and lasers[20–23]. The concept of supercoupling further opens the potentials for numerous novel photonic systems with distant energy transfer, such as supercoupled photonic chips, fiber optics, and cascade laser technologies.

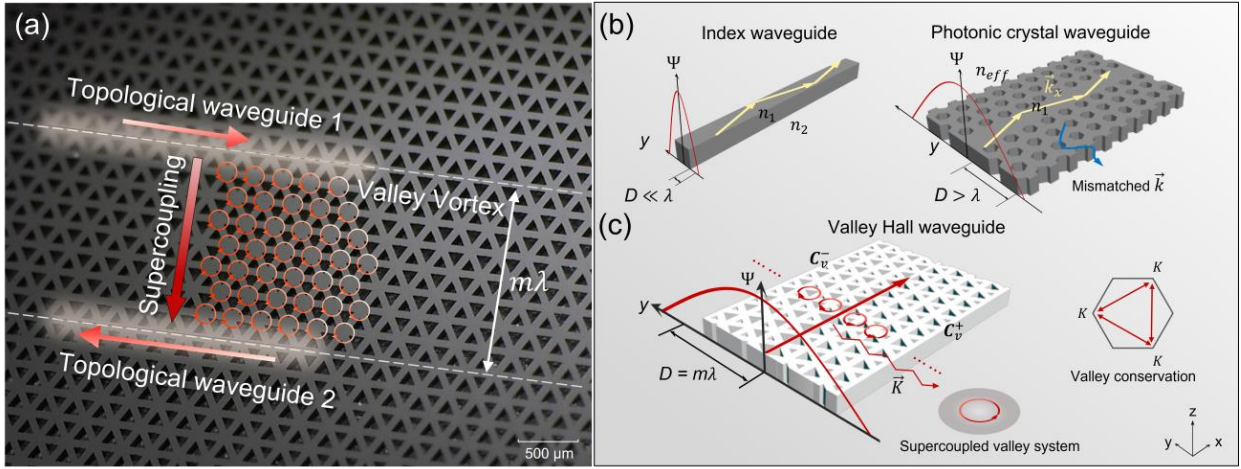

**FIG. 1**. Photonic supercoupling beyond multiple wavelengths in topological Valley-Hall system. (a) Image of waveguide supercoupling on a VPC chip. The topological waveguide mode is coupled into a secondary topological waveguide via valley-conserved vortices in the coupling region. A highly efficient energy transfer can be achieved albeit two parallel waveguides are separated $m\lambda$. (b) Schematic waveguide mode profiles ψ in index waveguide (left) and photonic crystal waveguide (right). The evanescent distances $D$ are determined by total internal reflection at the interface of media characterized by refractive indices $n_1$ and $n_2$ (or $n_{eff}$). Such evanescent tails lead to low coupling efficiency in long distance, challenged by manipulating evanescent length and addressing mismatched momentum. (c) Scheme of supercoupling in valley Hall waveguide with $D = m\lambda$. Interfaced by mirrored VPC bulks with inversed

valley Chern numbers $C_v^\pm$, the waveguide mode presents efficient intra-$K$-valley momentum transitions (inset of Brillouin zone) exhibiting vortex behavior and providing $y$-directional momenta. A supercoupled VPC system conserves the valley momentum and enables 100% energy transfer in a long distance.

**Results**

In conventional index waveguide system, as illustrated in the left panel of Fig. 1(b), the mode is confined by total internal reflection (TIR) at the boundary of materials with refractive indices $n_1$ and $n_2$, respectively. As a result, the mode amplitude (ψ) decays exponentially with an evanescent distance $D$ (where ψ is $1/e$ of maximum), that usually restricts the coupling distance to a narrow range ($D \ll \lambda$). Lengthening the evanescent tail into multiple wavelengths depends on the variation of the material index, that poses challenges in platforms with fixed permittivity. And achieving high coupling efficiency with such extended evanescent tail also requires an extremely long coupling length based on coupled mode theory. Photonic crystal waveguide can extend the coupling distance (right panel of Fig. 1(b)) by manipulating the periodic structure with effective refractive index ($n_{eff}$). Similarly, efficient waveguide mode coupling mostly occurs within λ to avoid the disordered momenta in the bulk, that can attenuate the coupling efficiency.

In contrast, the valley-conserved VPC system provides the potential for ultra-long-distance mode coupling. The VPC waveguide is the interface of two VPC bulks characterized by reversed valley Chern numbers $C_v^\pm$, achieved through breaking the spatial inversion symmetry. The topological waveguide mode (kink state[24]) exists at this interface within the topological bandgap as shown in Fig. 1(c), of which the mode profile is expressed as[25]:

$$\psi(\vec{r}) = |\psi_0|e^{-i\vec{K}_x \cdot \vec{x} \pm \alpha(\delta)|\vec{y}|}$$

where $\psi_0$ is the maximum mode profile amplitude at the interface ($y = 0$), $\vec{K}_x$ is the projected valley momentum along +$x$-direction, and $\alpha(\delta)$ is the decay rate along $\pm y$-direction as a function of the symmetry perturbation strength δ. The waveguide mode tail can be extended over several wavelengths ($D = m\lambda$) with a minimal δ (Fig. 1(c) and see Supporting Information (SI) S1). The $C_{3v}$ symmetric unit cells in VPC bulks conserve the valley momenta as featured by on-site valley vortex (red circles in Fig. 1(c)). This photonic vortex refers to the in-plane electromagnetic wave flow, resulting from the superimposition of all $K$-directional momenta. The symmetry-induced intra-$K$-valley conversion, as shown by the first Brillouin zone in the inset of Fig. 1(c), allows for sharp turning of the wave, circumventing defects, or attenuating backscattering. When introducing a coupled leaking valley-channel, these structured vortices facilitate the long-range energy transfer by providing $y$-directional momenta.

The waveguide supercoupling in VPC is schemed in Fig. 2(a), integrating two parallel topological interfaces in the shape of an A-B-A bulk sandwich configuration. The A and B bulks present opposite $C_v$, and the waveguides are indicated by WG1 (green) and WG2 (black). When the valley momentum $\vec{K}$ is injected into WG1 at $x = 0$, the whole system is valley-locked, and WG1 and WG2 present opposite group velocities. A contradirectional coupling[26,27] is achieved with pure valley momenta, that enables 100% energy transfer to WG2. Assuming lossless waveguiding, the waveguide supercoupling is expressed by the equations in $x > 0$ region:

$$\frac{d\psi_1}{dx} = -j\kappa_{sc}\psi_2 exp\left(j\left(\vec{K}_{x2} - \vec{K}_{x1} + \sum_i \vec{K}_i\right) \cdot \vec{x}\right)$$

$$\frac{d\psi_2}{dx} = j\kappa_{sc}\psi_1 exp\left(-j\left(\vec{K}_{x2} - \vec{K}_{x1} + \sum_i \vec{K}_i\right) \cdot \vec{x}\right)$$

The momentum matching between WG1 and WG2, with respective $x$-projected valley momenta $\vec{K}_{x1}$ and $\vec{K}_{x2}$, is achieved via valley-conserved coupling area where the effective momentum is $\Sigma\vec{K}_i$, $i$ indicates each unit cell site. The efficient intra-$K$-valley momentum transition gives rise to the rotation-invariant wave flows forming vortex array over the coupling region. The projection angle $\theta_i$ between $\vec{K}_i$ and $\hat{e}_x$ (Fig. 2(a)) can be thus expressed as $\theta_i = \theta_0 + q_i \cdot 2\pi/3$ where $\theta_0$ is $\pm\pi/3$ for parallel waveguides, $q_i = 1, 2, 3$. The valley-conserved momenta lead to a largely-enhanced supercoupling coefficient $\kappa_{sc}$ enabling 100% energy transfer over $m\lambda$ distance and a significantly short coupling length. The reciprocal coupling coefficient also enables the backward complete coupling from WG2 to WG1 at $x < 0$ region.

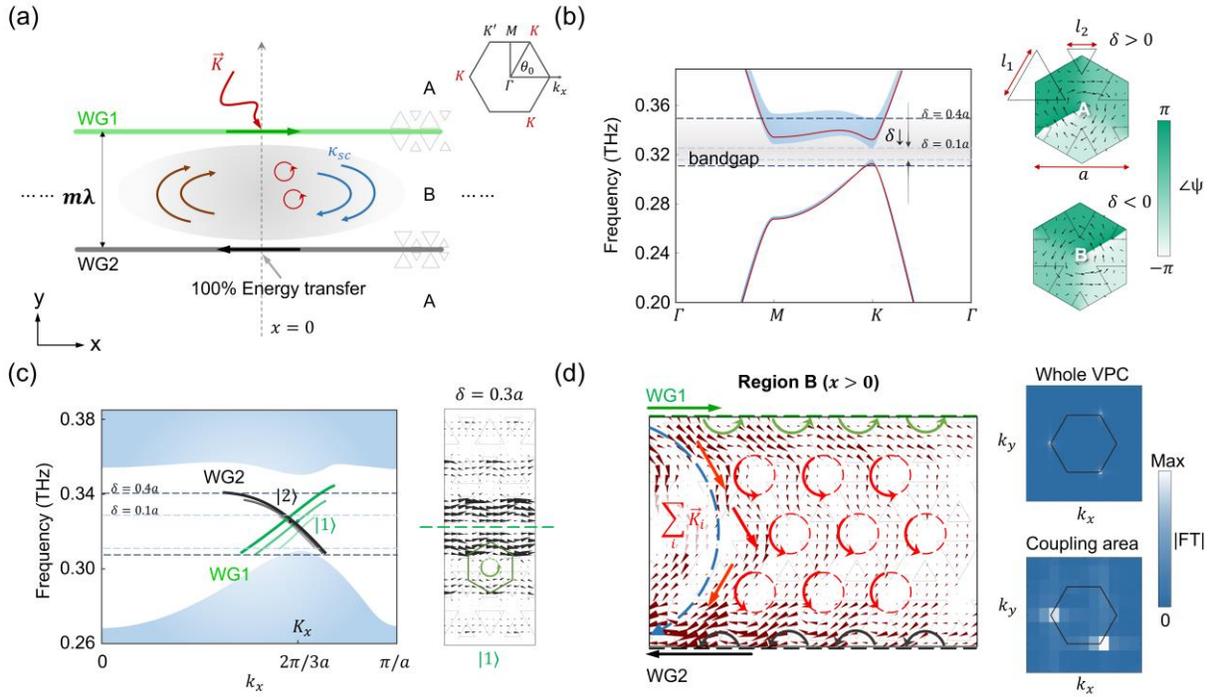

**FIG. 2**. Topological waveguide supercoupling based on in-plane valley-conserved vortex. (a) The schematic of contra-directional parallel waveguide supercoupling. Interfaces by A-B-A bulks with opposite topological invariants, two topological waveguides (WG1 in green and WG2 in black) present opposite group velocity when the system is $K$-valley-locked. The valley vortex array forming supercoupling loops mediates the coupling region for long distant ($m\lambda$) 100% energy transfer. (b) Valley vortex and $\delta$-dependent topological bandgap. Left: The topological photonic band diagram of VPC. The topological bandgap is lifted at $K$ valley which is narrowed with decreased $\delta$ (grey region). Right: The eigenmodes exhibiting vortex behavior (Poynting vector in black arrows) with phase singularities ($\angle\psi$ in colormap) of unit cell A (B) with $\delta > 0$ ($\delta < 0$). (c) Left: Band structures of topological kink states for WG1 (green) and WG2 (black), characterized by opposite group velocities with tuned bandwidth versus $\delta$. Right: The electromagnetic flow of state $|1\rangle$ ($\delta = 0.3a$) composed of partial valley-vortices (green circle) along $x$-direction. (d) Zoom-in valley-conserved wave flow in the coupling region ($x > 0$ in (a)) with the Fourier transform results of the whole system and the coupling region. $|1\rangle$ (green arrow), vortex-based wave in coupling region (red circles and arrows), and $|2\rangle$ (black arrow) form supercoupling loops (blue dash) as the valley-conserved wave transfer channels.

The valley vortex is illustrated by the simulated unit cell eigenmode in the right panel of Fig. 2(b). The symmetry breaking is achieved by altering the on-site equilateral triangular hole sizes from $l_1 = l_2$ to $l_1 \neq l_2$, where $l_1$ and $l_2$ represent the side lengths of the triangles, and $a$ is the lattice constant. The symmetry perturbation strength is defined as $\delta = l_1 - l_2$. The transverse-electric valley vortex is showcased for type A ($\delta > 0$) and type B ($\delta < 0$) unit cells by the simulated topological phase singularities ($\angle\psi$ in colormap, $\psi$ indicates magnetic field) and Poynting vector vortex ($\vec{S}$ in arrows). The simulated photonic band diagrams of the VPC unit cells are presented in Fig. 2(b) (left panel) where $\delta$ is from $0.1a$ to $0.4a$ (blue shadow), leading to a tunable topological bandgap[28]. The corresponding kink states also present $\delta$-dependent bandwidth, as shown on the left of Fig. 2(c), where a narrowly gapped kink state gives rise to an extended mode tail. Partial vortices exist near the interface, as presented by the simulated $\vec{S}$ (black arrow) of $|1\rangle$ ($\delta = 0.3a$) of WG1 in the right panel of Fig. 2(c). Note that while pure topological waveguiding is immune to perpendicularly leaking, the intra-valley momentum transition occurs when introducing a secondary valley coupled channel.

The valley-conserved vortex array in the coupling region is exhibited in Fig. 2(d) with the simulated $\vec{S}$ (red arrows) in $x > 0$ region. When the source waveguide is excited by state $|1\rangle$ with partial vortices (green arrows), the whole system is $K$-valley conserved, and the chirality of region B is locked as counter-clockwise. The vortex flow in coupling region also presents efficient intra-$K$ transition while holding immunity of $K - K'$ valley scattering, as illustrated by the Fourier transform results (right panel of Fig. 2(d)). These vortex-based wave flows with effective momentum $\Sigma\vec{K}_i$, as well as the states $|1\rangle$ and $|2\rangle$ (black arrows), collectively form a series of supercoupling loops (blue dash) as wave transferring channels. The presences of supercoupling loops with enhanced $\vec{K} \cdot \hat{e}_y$ enable efficient energy accumulation in WG2, allowing for complete wave transfer even over multiple wavelengths.

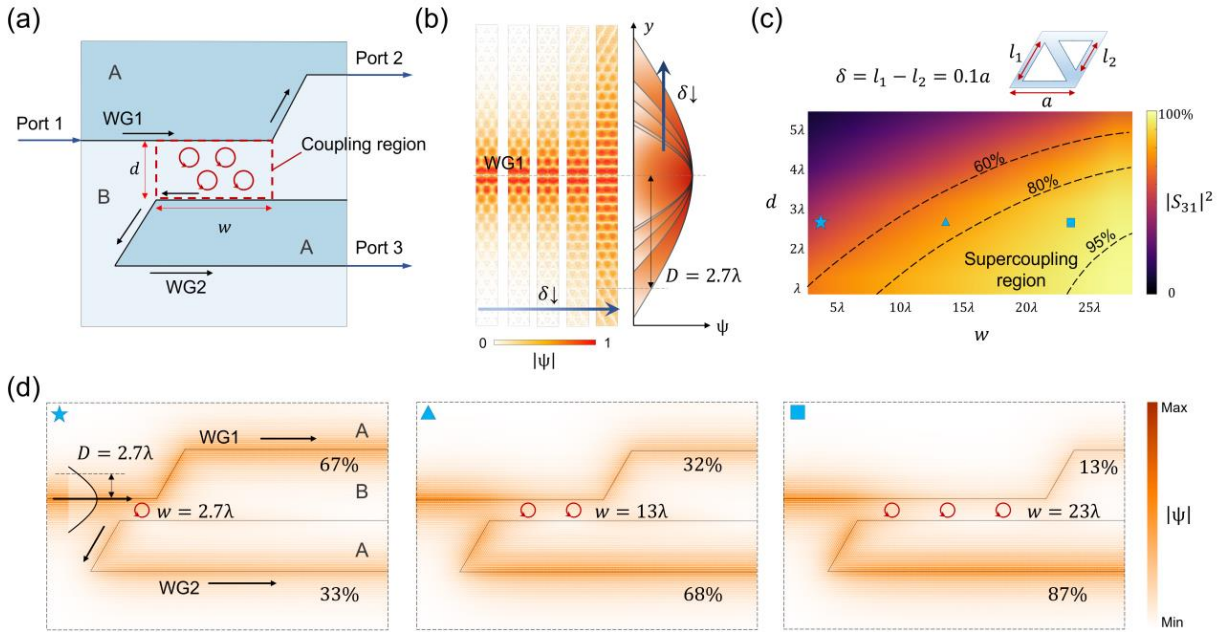

**FIG. 3**. Supercoupling at long separation distance with high coupling ratio. (a) The scheme of supercoupling through rectangular coupling region (red box) with distance $d$ and length $w$. The coupling ratio is represented by the powers transmitted from port 1 to port 2 and port 3, respectively. (b) The waveguide mode profiles against δ. Eigenmode tail length $D$ increases with decreasing δ, as presented by the normalized simulated ψ of the example interfaces. $D$ is extended into 2.7λ when δ = 0.1$a$. (c) Transmittance $|S_{31}|^2$ against $d$ and $w$ when δ = 0.1$a$ at center frequency. An over-80% coupling effect at $d$ = 4λ can be achieved with certain length $w$ and over 95% coupling ratio can be achieved even when $d$ > 2λ. (d) Simulated wave propagating profiles when $d$ = 2.7λ and $w$ = 2.7λ (star), 13λ (triangle), and 23λ (square), where the coupling ratios are 33%, 68%, and 87%, respectively with sufficient supercoupling loops.

To assess the coupling distance limit, we constructed a rectangular coupling region (red dashed box) as depicted in Fig. 3(a). The coupling ratio of two parallel topological waveguides depends on the efficiency of each supercoupling loop and the number of loops implemented. These parameters are characterized by the coupling distance $d$ and the coupling length $w$, respectively. The robust VPC waveguides are bent into a sharp angle to guide the wave rightward with minimized bending loss[24]. When the wave is fed from port 1, the coupling effect can be monitored by observing the output powers at port 2 and port 3. δ is chosen as (0.55-0.45)$a$ = 0.1$a$ with an extended mode tail as $D$ = 2.7λ, shown by the variation trend of simulated eigen mode profiles in Fig. 3(b) (also see SI S1). Figure 3(c) illustrates the simulated power S-parameter $|S_{31}|^2$ (colormap) with tuned $d$ and $w$ at center frequency. An over 95% $|S_{31}|^2$ (region below 95% dash line) can be realized albeit $d$ > 2λ. These results indicate an ultra-highly efficient long-distant coupling process, significantly exceeding that of the evanescent waveguide coupling, as compared in SI S2. The $w$-dependent coupling results are showcased by three simulated electromagnetic wave flows in Fig. 3(d) (ψ in colormap). For a coupling distance of $d$ = 2.7λ, there is a drastic increase in energy coupled from WG1 to WG2 accounting for the growing number of valley-conserved supercoupling loops, as $w$ varies from 2.7λ (star) to 13λ (triangle), and to 23λ (square). The coupled $|S_{31}|^2$ values are 33%, 68%, and remarkably 87% respectively. The respective coupling strength $\kappa_{sc}w$ are 0.65, 1.17 and 1.68, obtained using equation $|S_{31}|^2 = \tanh^2(\kappa_{sc}w)$ [27]. Thus, the simulated $\kappa_{sc}$ is equal to $0.05\lambda^{-1}$ when $d$ = 2.7λ. Additional coupling results with varying δ are presented in SI S3.

Experimentally, the topological waveguide supercoupling configuration is fabricated on a 200-μm-thick high-resistivity silicon chip in the terahertz domain, as imaged in Fig. 4(a). The VPC has a period of $a$ = 242.5 μm and perturbation of δ = 0.1$a$, forming a topological band near the center frequency of 0.33 THz. The respective lengths of hole edge are $l_1$ = 133 μm (0.55$a$) and $l_2$ = 109 μm (0.45$a$). The terahertz wave is generated utilizing a frequency multiplier (Virginia diode inc., see SI S4). The wave is coupled into and out of the waveguide through WR-2.8 hollow waveguides and the silicon tapered coupler attached to the chip. The valley-vortex enables an efficient adiabatically mode transition from plane wave to state |1⟩. The on-chip supercoupling ratio is represented by the ratio of transmissions through WG1 and WG2 as: $R = S_{31}/(S_{21} + S_{31})$, where $R$ is the coupling ratio, and $S_{21}$ and $S_{31}$ indicate the respective transmissions. The maximum coupling ratios of four devices under test (with $d$ = 1.8λ, 2.7λ, 3.6λ, and 5λ when $w$ = 26λ) are illustrated in the top panel of Fig. 4(b) with both simulated (red) and experimental (blue) results. Near the central frequency, the coupling ratios reach their peak values. Both the simulated and experimental results demonstrate $R$ values exceeding 95% when the coupling distance $d$ is either 1.8λ or 2.7λ. When $d$ is increased to even 5λ, the simulation still shows a coupling ratio of 69%, while the experimental result decreases to 31%. The corresponding max $S_{31}$ is also presented in the bottom panel of Fig. 4(b), where the simulated $S_{31}$ reaches the level of 0.7 and the measured results are from 0.46 for $d$ = 2.7λ to 0.17 for $d$ = 5λ. This experimental transmission loss is attributed to the intrinsic material loss, particularly for a relatively lengthy on-chip propagation path spanning up to 5 cm. And it is important to note that further extending such long coupling distances is possible through tailoring geometric parameters $w$ to introduce additional supercoupling loops, or by adjusting the symmetry perturbation parameter $\delta$ to increase the mode tail distance $D$.

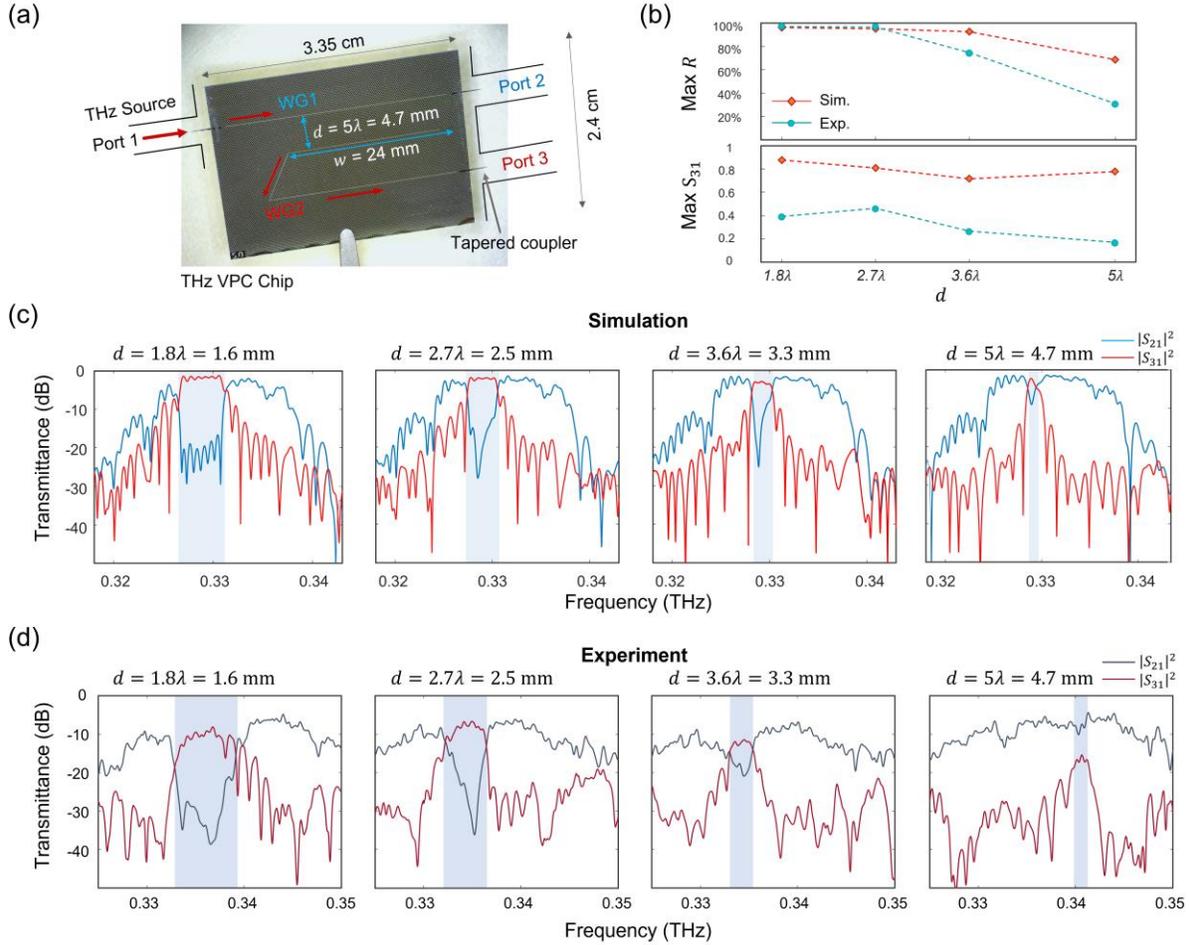

**FIG.4**. Experimental demonstration of waveguide supercoupling on a terahertz VPC chip. (a) Image of waveguide supercoupling chip. The terahertz wave is fed into the chip via port 1 and is received from port 2 and port 3. The topological waveguide coupling ratio is represented by ratio of transmittances through WG1 and WG2, respectively. The on-chip coupling distance can reach 4.7 mm (5λ). (b) The simulated and experimental results of maximum coupling ratio $R$ (top) and $S_{31}$ (bottom) when $d$ is 1.8λ, 2.7λ, 3.6λ, and 5λ, and $w = 26λ$. (c,d) The comparison of simulated and experimental transmittance spectra of $|S_{21}|^2$ and $|S_{31}|^2$ for $d$ = 1.8λ, 2.7λ, 3.6λ, and 5λ, respectively.

The simulated and measured on-chip terahertz transmittance spectra are illustrated in Fig. 4(c) and 4(d) for all the four tested chips, respectively. In Fig. 4(c), the simulated results show high transmittances of WG2 in supercoupling bandwidth (blue regions) even when $d$ increases into 5λ. In Fig. 4(d), although minor frequency shifts (< 0.01 THz) are observed when compared to the simulated results, there remains a high degree of consistency between simulated and experimental results. The frequency shift results from the fluctuations of chip thickness and fabrication accuracy, and the experimental findings indicate a power loss of 5 dB in both $|S_{21}|^2$ and $|S_{31}|^2$.

The waveguide supercoupling exhibits dispersive characteristics since the topological $K$-valley vortex is maximally enhanced at the center frequency. The coupling ratio reaches maximum at center frequency and diminishes as the frequency diverges from this point (see SI S5). The supercoupling process also exhibits a robustness property based on valley conservation when coupling into non-parallel directions (see SI S6) or when introducing defects in coupling region (see SI S7). Possible practical implementations of the supercoupling on an integrated chip can be achieved through locally adjusting δ in a tapered waveguide[29] or by incorporating waveguide modes with asymmetric $D$ in each side of the interface[30]. These engineering techniques enrich the on-chip supercoupling designs for innovative supercoupled photonic devices, improving the spatial and energy efficiency.

**Discussion**

In conclusion, we have demonstrated the supercoupling phenomenon through a terahertz waveguide coupling experiment in a topological valley Hall system. Leveraging the expansive mode extension and the valley-conserved vortex array, we significantly enhance waveguide coupling distances with an ultra-high coupling ratio. Our experiments validate an efficient coupling distance of 2.7 wavelengths with a coupling ratio of 95%, and even over 5 wavelengths with a reduced coupling ratio of 31%. Photonic supercoupling expands the efficiency boundaries of conventional evanescent coupling, unlocking possibilities for advanced supercoupled photonic systems such as topology-protected interferometers, integrated lasers, and topological-chip-based polaritonics. Supercoupling also opens doors to emerging fields including on-chip supercoupled sensors, couplers, and spectrometers.

**Acknowledgement**

This work is supported by National Research Foundation (NRF) Singapore, Grant No: NRF-CRP23-2019-0005.